\documentclass[11pt,twoside]{article}

\usepackage{asp2006}
\usepackage{epsf}
\usepackage{psfig}
\usepackage{lscape}

\usepackage{natbib}

\begin{document}
\title{A Large Population of High Redshift Galaxy Clusters in the IRAC Shallow Cluster Survey}   
\author{Mark Brodwin\altaffilmark{1}, Peter R.~Eisenhardt\altaffilmark{2}, Anthony H.~Gonzalez\altaffilmark{3}, S.~Adam Stanford\altaffilmark{4}, Daniel Stern\altaffilmark{2}, Leonidas A.~Moustakas\altaffilmark{2}, Michael J.~I.~Brown\altaffilmark{5}, Ranga-Ram Chary\altaffilmark{6}, Audrey Galametz\altaffilmark{2,7}}   
\altaffiltext{1}{NOAO, 950 N.~Cherry Ave., Tucson, AZ 85719}
\altaffiltext{2}{JPL/Caltech, 4800 Oak Grove Dr., MS 169--327, Pasadena, CA 91109}
\altaffiltext{3}{Dept.~of Astronomy, University of Florida, Gainesville, FL 32611}
\altaffiltext{4}{University of California, Davis, CA 95616}
\altaffiltext{5}{School of Physics, Monash University, Clayton, Victoria 3800, Australia}
\altaffiltext{6}{SSC, California Institute of Technology, Mail Stop 220-6, Pasadena, CA 91125}
\altaffiltext{7}{Observatoire Astronomique de Strasbourg, 11 rue de l'Universit\'e, 67000 Strasbourg, France}


\begin{abstract} 

  We have identified 335 galaxy cluster and group candidates spanning
  $0<z<2$, using a 4.5$\mu$m selected sample of galaxies in a 7.25
  deg$^2$ region in the {\it Spitzer}/IRAC Shallow Survey. Using full
  redshift probability distributions for all galaxies, clusters were
  identified as 3-dimensional overdensities using a wavelet algorithm.
  To date 12 clusters at $z>1$, and over 60 at $z < 0.5$ have been
  spectroscopically confirmed.  The mean I-[3.6] color for cluster
  galaxies up to $z\sim 1$ is well matched by a $z_f=3$ passively
  evolving model.  At $z>1$, a wider range of formation histories is
  needed, but higher formation redshifts (i.e.\ $z_f \ge 4-5$) are
  favored for most clusters.  The cluster autocorrelation function,
  measured for the first time out to $z=1.5$, is found not to have
  evolved over the last 10 Gyr, in agreement with the prediction from
  $\Lambda$CDM.  The average mass of the IRAC Shallow Cluster Survey
  sample, inferred from its clustering, is $\sim 10^{14} M_\odot$.
\end{abstract}


\section{Introduction}

The bulk of the stellar mass in the universe is created at $1<z<3$
\citep{dickinson03, rudnick06} in very massive luminous and
ultraluminous galaxies \citep[LIRGs and ULIRGs,][]{lefloch05,
  papovich06}.  Recent studies show that the SFR at $z=1$ increases
with increasing local density \citep{elbaz07, cooper08}, a reversal of
the local SFR-density relation \citep{lewis02, brinchmann04}.  To
obtain a complete census of massive galaxy formation at $z>1$ we need
to study it across a range of environments, not least because most
massive galaxies are found in very rich environments.  Here we
describe the IRAC Shallow Cluster Survey \citep[ISCS;][hereafter
E08]{eisenhardt08}, the largest, best-studied sample of stellar mass
selected clusters at $z>1$.

\section{The IRAC Shallow Cluster Survey}

\begin{figure}[bthp]
\vspace*{-0.25cm}
\plotfiddle{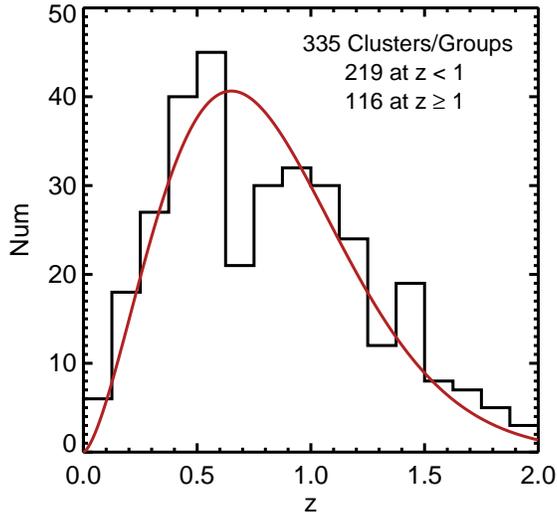}{3in}{0.}{40.}{40.}{-100.}{0}
\vspace*{-0.5cm}
\caption{Observed redshift distribution of 335 galaxy clusters in the
  ISCS.  The curve, a fit to the distribution, is used in the
  clustering analysis in \textsection{4}.}
\label{N(z)}
\end{figure}

The ISCS is a sample of 335 galaxy clusters spanning $0.1<z<2$ in the
IRAC Shallow Survey \citep[ISS,][]{eisenhardt04} area of the NOAO
Deep-Wide Field Survey (NDWFS, \citealt{ndwfs99}; Jannuzi et al.\ in
prep) in Bo\"otes.  The clusters are selected using a wavelet
detection algorithm which identifies peaks in cluster probability
density maps constructed from accurate photometric redshift
probability functions for 175,431 galaxies brighter than 13.3$\mu$Jy
at 4.5$\mu$m in a 7.25 deg$^2$ region (\citealt{brodwin06_ISS},
hereafter B06, E08).  The redshift distribution is shown in Fig.\
\ref{N(z)}.

The AGES survey (Kochanek et al.\ in prep) in Bo\"otes provides
spectroscopic confirmation for over 60 clusters at $z\le 0.5$ (E08).
At higher redshift, a multi--year Keck spectroscopic campaign has to
date confirmed 12 $z>1$ clusters, many with $\sim 10-20$ members, at
redshifts from 1.06 to 1.41 (\citealt{stanford05}; \citealt{elston06};
B06; E08).

\section{Formation of Massive Galaxies}

\begin{figure}[bthp]
\plottwo{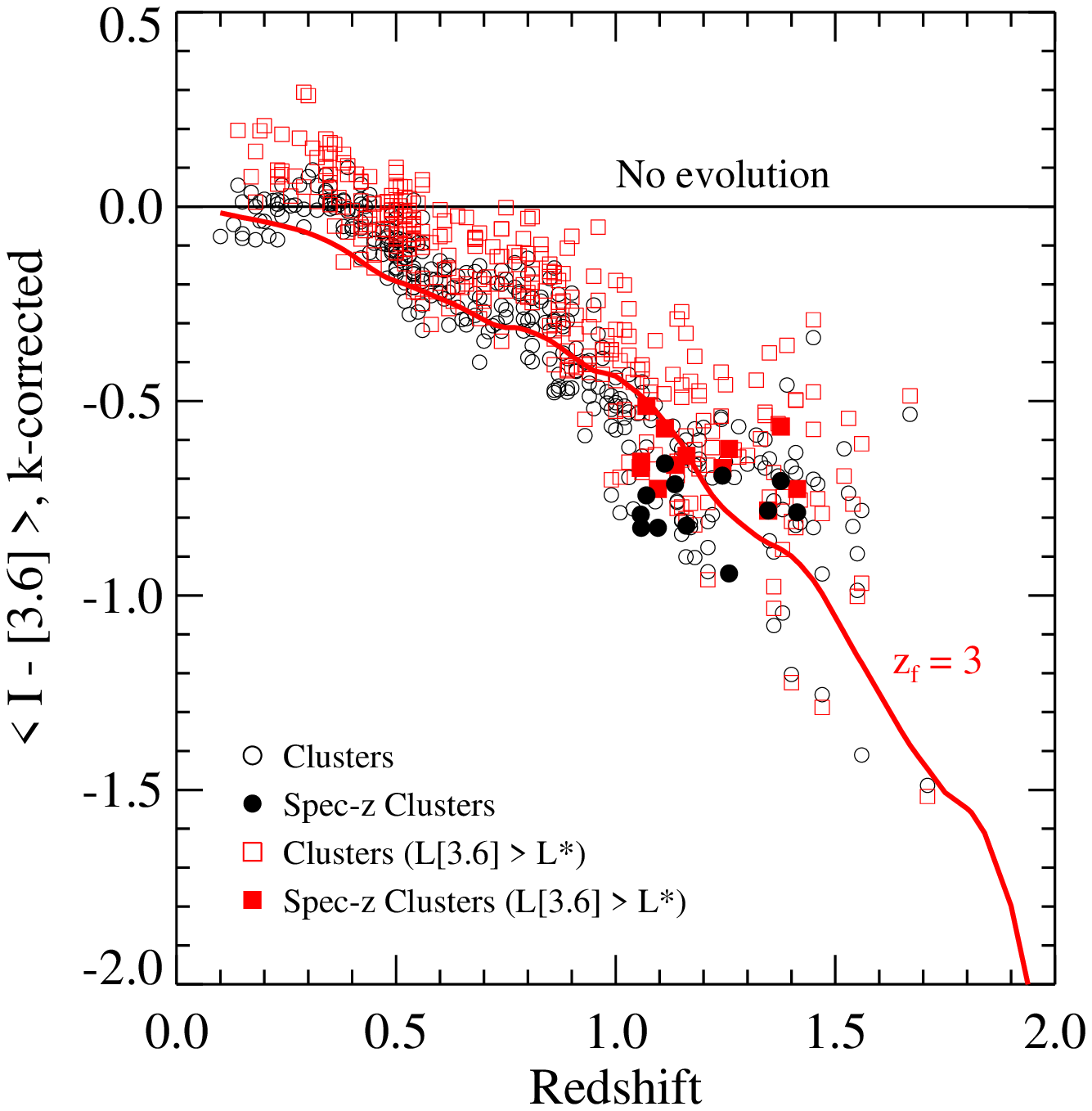}{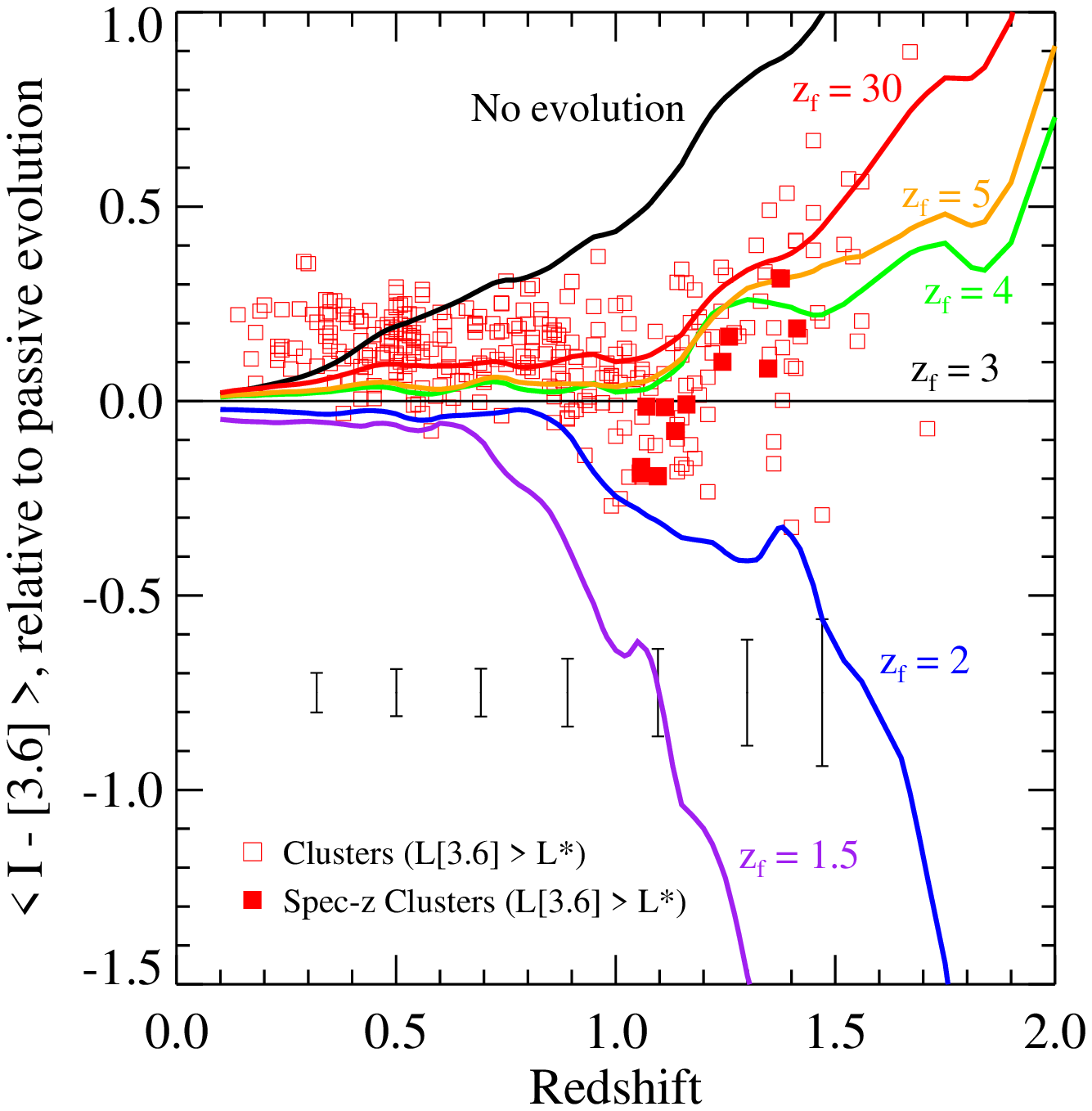}
\caption{({\it left)}: Mean, k-corrected I-[3.6] color vs.\ redshift
  for clusters in the ISCS (circles; filled symbols are confirmed
  $z>1$ clusters).  The red curve is a passively evolving $z_f=3$
  model.  Red boxes are the mean colors of massive ($L>L^*$) galaxies
  only and are systematically redder than the overall cluster
  population. {\it right}: Mean I-[3.6] color relative to the $z_f=3$
  PE model.  Clusters at $z\ga 1.2$ formed their stars at very high
  redshift ($z_f \ga 4$). }
\label{SFH}
\end{figure}

The extensive redshift baseline of the ISCS sample permits a test of
various plausible star formation histories (E08).  Fig.\ \ref{SFH}
({\it left}) plots the average k-corrected I-[3.6] color for galaxies
within 1 Mpc of each cluster center whose integrated redshift
probability distribution in the range $z_{\mbox{\scriptsize cl}} \pm
0.06(1+z_{\mbox{\scriptsize cl}})$ exceeds 0.3 (circles; filled
symbols indicate $z>1$ spectroscopically confirmed clusters.  The red
curve is not a fit, but rather a \citet{bc03} population synthesis
model in which the stars formed in a 0.1 Gyr burst at $z_f=3$ and
evolved passively thereafter.  This scenario clearly describes the
data better than a no-evolution model.  Also plotted are the average
k-corrected colors for galaxies more massive than a passively evolving
$L^*$ galaxy (red boxes).  These are systematically redder than the
full cluster population, showing the persistence of the
color-magnitude relation in clusters out to $z\approx 1.5$, and
suggesting the early existence of a mass-metallicity relation in
clusters and/or providing evidence of downsizing.

In Fig.\ \ref{SFH} ({\it right}) the average k-corrected color of $L >
L^*$ cluster galaxies is plotted relative to the $z_f=3$ passive
evolution model (horizontal line).  Overplotted are models with
formation redshifts of $z_f =[1.5, 2, 4, 5, 30]$, as well as a
no-evolution model.  The offset between all models and the data at low
redshift is primarily due to the redder colors of these massive
galaxies discussed above.  Nevertheless we can draw two robust
conclusions from this figure regarding the star formation histories of
$z>1$ clusters.  First, the variation in the average cluster colors
increases markedly at $z>1$.  Second, we find that the stars in most
high-redshift clusters, particularly those above $z>1.2$, formed at $z
\sim 4-5$ or even earlier.

\section{Clustering of Galaxy Clusters to $z=1.5$}

The large, uniformly selected ISCS cluster sample has permitted the
first measurement of the galaxy cluster autocorrelation function in
the first half of the Universe \citep{brodwin07}.  Specifically, the
autocorrelation function was measured in two broad bins, at $0.25 \le
z \le 0.75$ and $0.75 < z \le 1.5$, with resulting clustering
amplitudes of $r_0 = 17.40^{+3.98}_{-3.10}$ and
$r_0=19.14^{+5.65}_{-4.56}$ h$^{-1}$ Mpc, corresponding to average
halo masses of $\log [M_{\mbox{\scriptsize 200}}/M_\odot]$ $\sim
13.9^{+0.3}_{-0.2}$ and $\sim 13.8^{+0.2}_{-0.3}$, respectively.
Errors are computed via bootstrap resampling of the cluster catalog
and are therefore rather conservative.

\begin{figure}[bthp]
  \plottwo{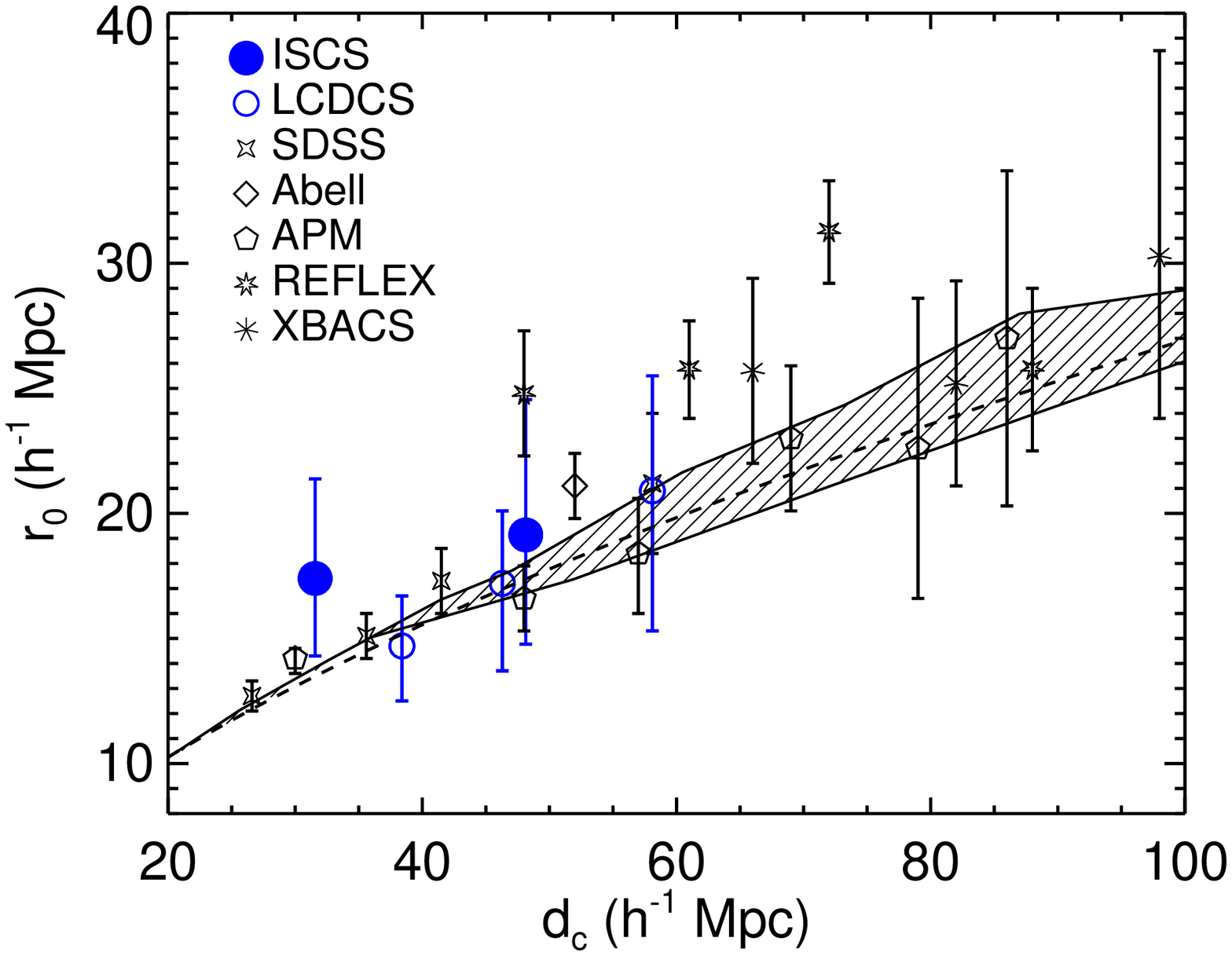}{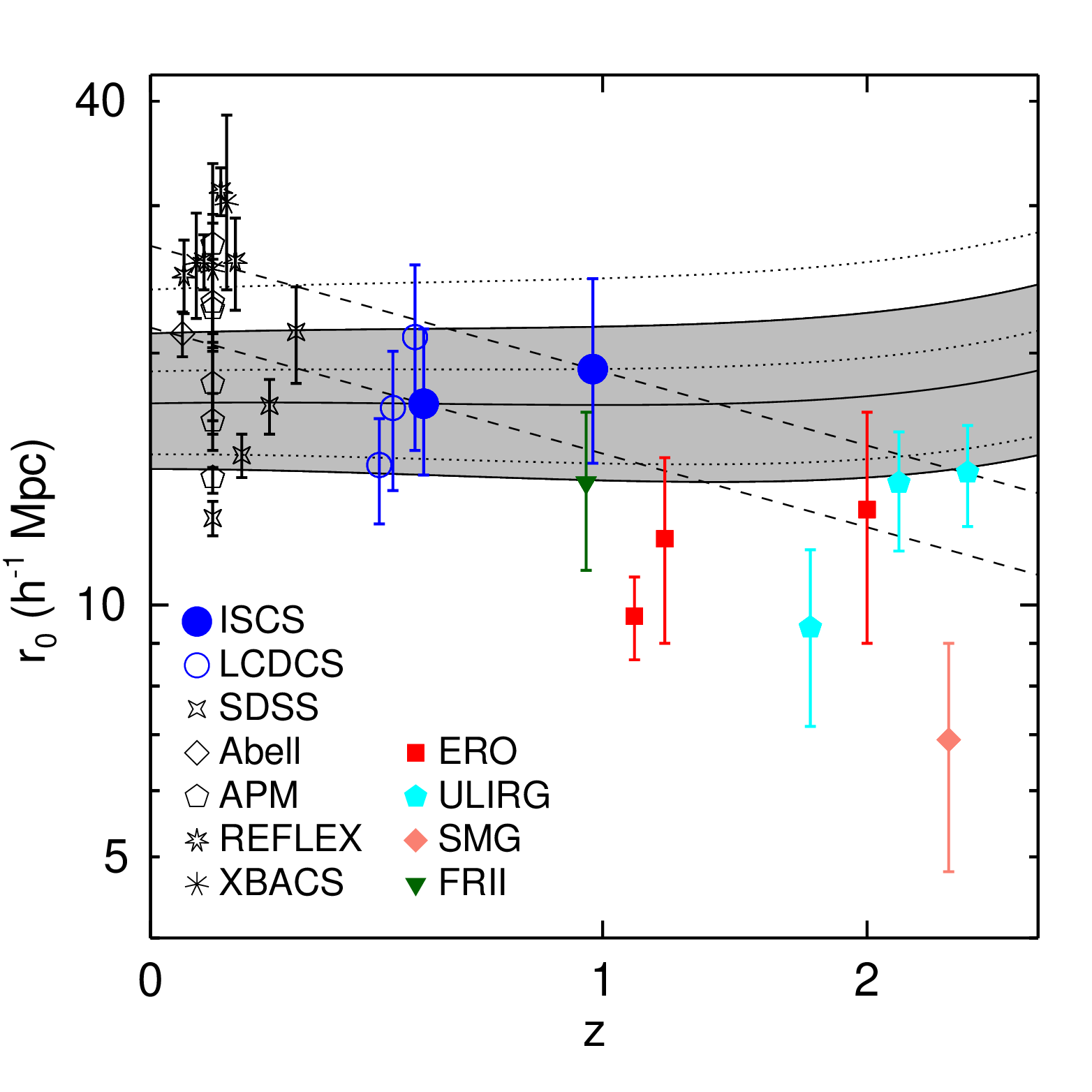}
  \caption{{\it Left}: $r_0$ vs.\ $d_c$ for the present sample (filled
    circles) along with several $z\la 0.5$ measurements from the
    literature \citep[][and references therein]{bahcall03}.  The
    hashed region shows the $\Lambda$CDM prediction over $0 <z<1.5$
    \citep{younger05}.  {\it Right}: Comoving correlation lengths for
    the ISCS clusters, other cluster samples at low-z, and high-z,
    highly clustered galaxy populations. The filled region (dotted
    lines) illustrates the predicted $r_0$ evolution in a biased
    structure formation model for our $z_{\mbox{\scriptsize
        eff}}=0.53$ ($0.97$) cluster samples.  The dashed lines
    represent simple stable clustering models.}
\label{r0}
\end{figure}

These correlation lengths are quite similar to the values for local
clusters (e.g.\ Abell, APM, SDSS) as well as those at $z \approx 0.5$
\citep{gonzalez02}.  This constancy with redshift is a key prediction
of $\Lambda$CDM, as shown in Fig.\ \ref{r0} ({\it left}).
Measurements of the cluster clustering amplitude, $r_0$, are plotted
vs.\ intercluster distance, $d_c$, for several different surveys
spanning redshifts up to $z\approx 0.5$.  Overplotted are the present
measurements, at effective redshifts of 0.5 and 1.0 (filled symbols).
The hashed regions shows the prediction from a high-resolution
numerical simulation \citep{younger05} that over $0 <z< 1.5$ very
little evolution is expected, in agreement with our results.

In Figure \ref{r0} ({\it right}) we plot a compilation of recent
clustering amplitudes for various cluster surveys, as well as for
highly clustered galaxy populations.  Following
\citet{moustakas&somerville02}, we overplot the halo conserving model
of \citet{fry96} normalized to our two measurements in order to
explore possible evolutionary connections with structures at other
redshifts.  The shaded area (dotted lines) shows the 1\,$\sigma$
region for the $z_{\mbox{\scriptsize eff}}=0.53$ ($0.97$) measurement.
In this model the ISCS clusters will evolve into typical present--day
massive clusters, such as those in the SDSS, APM or Abell surveys.  In
the stable--clustering picture, in which clustering is fixed in
physical coordinates \citep[dashed lines]{groth&peebles77}, the
$z_{\mbox{\scriptsize eff}}=0.97$ ISCS clusters grow into the most
massive clusters in the local Universe, typically identified in X--ray
surveys.

Most of the plotted high redshift galaxy clustering measurements are
rather uncertain due to both small number statistics and poorly known
redshift distributions.  Several authors also find clustering
amplitudes for galaxy populations which are inconsistent with their
space densities \citep[e.g.][]{quadri07}.  Indeed, only obscured
ULIRGs have space densities similar to the present cluster sample, a
prerequisite for drawing evolutionary connections from these
particular models.  Their clustering is consistent with that of the
ISCS clusters, providing a measure of support for recent studies
\citep[e.g.][]{magliocchetti07} indicating that they may be associated
with, or progenitors of, groups or low--mass clusters.

\acknowledgements 
We'd like to thank the organizers for organizing an excellent
conference and for awesome sushi at the banquet.  This work is based
in part on observations made with the Spitzer Space Telescope, which
is operated by the Jet Propulsion Laboratory, California Institute of
Technology under a contract with NASA. This paper made use of data
from the NDWFS, which was supported by NOAO, AURA, Inc., and the
NSF. We thank A.~Dey, B.~Jannuzi and the entire NDWFS survey team.
NOAO is operated by AURA, Inc., under a cooperative agreement with the
National Science Foundation.
\bibliographystyle{astron3}
\bibliography{bibfile}

\end{document}